\documentclass[journal]{IEEEtran}
\usepackage{siunitx}
\usepackage{mathrsfs}
\usepackage{color}
\usepackage{enumerate,graphicx,epstopdf}
\usepackage{amsmath,amssymb,bm}
\usepackage{cite}
\usepackage{mathtools}
\usepackage{array}
\usepackage{algpseudocode,algorithm,algorithmicx}
\usepackage{booktabs}
\usepackage[hyphens]{url}
\usepackage{lipsum}
\usepackage{graphicx}
\usepackage[deletedmarkup=souauthormarkup=superscript]{changes}
\usepackage[normalem]{ulem}
\usepackage{flushend}
\begin{document}

\title{Loss aware pricing strategies for peer to peer energy trading}

\author{Varsha N. Behrunani$^{1,2}$, Philipp Heer$^{2}$, Roy S. Smith$^{1}$ and John Lygeros$^{1}$
\thanks{This research is supported by the SNSF through NCCR Automation (Grant Number 180545). The authoes also want to thank Dr. Giuseppe Belgioioso for his insightful comments during this work.}
\thanks{$^{1}$V. Behrunani, J. Lygeros and Roy S. Smith are with the Automatic Control Laboratory, ETH Zurich, Switzerland (email: bvarsha@ethz.ch, jlygeros@ethz.ch, rsmith@ethz.ch)}
\thanks{$^{2}$V. Behrunani and P. Heer are with the Urban Energy Eystems Laboratory, Swiss Federal Institute of Material Science and Technology (Empa) Dubendorf, Switzerland.(email: varsha.behrunani@empa.ch, philipp.heer@empa.ch)}
}

\maketitle

\begin{abstract}

Peer-to-peer(P2P) energy trading may increase efficiency and reduce costs, but introduces significant challenges for network operators such as maintaining grid reliability, accounting for network losses, and redistributing costs equitably. We propose a novel loss-aware pricing strategy for P2P energy markets that addresses these challenges while incentivizing participation in the cooperative energy trading market. The problem is formulated as a hierarchical Stackelberg game, where a grid operator determines network tariffs while prosumers optimize their trades based on these tariffs while guaranteeing that network constraints are satisfied. The algorithm is designed to minimize and recover their cost from the trading parties, while also minimizing the total cost of the hubs. The mechanism dynamically adjusts tariffs based on location and network topology, discouraging loss-intensive trades. Finally, the complete framework includes the computation of fair trading prices, ensuring all market participants benefit equitably. An ADMM-based hyper-gradient descent method is proposed for solving this problem. Extensive numerical simulations using the benchmark IEEE 33-bus system demonstrate significant cost reductions and improved network efficiency through reduction in network losses compared to constant tariff schemes. Results highlight the adaptability and scalability of the proposed mechanism to varying network configurations and size, demand profiles, and seasonal conditions.
\end{abstract}
\vspace{-0.1cm}

\IEEEpeerreviewmaketitle
\vspace{-0.3cm}
\section{Introduction}

The rising global energy demand coupled with increasing environmental and sustainability related concerns has led to an increase in renewable energy sources and developments in multi-generation and storage technology. This, in turn had left to a shift of the energy landscape from the traditional centralized system to a new decentralized network structure comprising multi-energy hubs and prosumers that act as both consumers and producers that are integrated into established systems~\cite{BOUFFARD20084504}. Energy hubs act as central nodes that optimize the production, conversion, storage, and distribution of multiple energy carriers to meet demands~\cite{Geidl:2006a}. They promote local energy generation and consumption, reducing reliance on centralized grids and promoting flexibility~\cite{Smith:2022, DARVISHI2024113624}.

This paradigm shift has also paved the way for the active participation of energy hubs in local peer-to-peer (P2P) markets. In addition to further reducing the overall energy cost and energy imports of the hubs~\cite{almassalkhi:2011z ,BEHRUNANI2024105922}, P2P trading also has the potential to reduce peak demand and reserve requirements~\cite{10591316}. The increased focus on energy resilience has made P2P trading a viable and appealing solution, particularly in remote or underserved areas. Additionally, the ability to monetize surplus energy and engage in community-focused energy exchanges further attracts prosumers and encourages the introduction of P2P trading mechanisms~\cite{GEORGARAKIS2021112615}. Technological advancement in trading mechanisms such as blockchain and smart contracts that facilitate easy, secure, low-cost automated transactions, and build trust in trading platforms has fast-tracked this transition~\cite{ZEDAN2024100778}. Finally, a key factor in P2P energy trading markets gaining traction globally is regulatory support that has been increasing all over the world as more governments are supporting P2P markets through pilot programs and legal frameworks\cite{Winssolutions_2024,shaw_2024}. In Switzerland specifically, a law encouraging the formation of local electricity communities, enabling collective self-consumption and facilitating P2P (P2P) energy trading among prosumers was enacted in 2018 and extended in January 2025~\cite{Deboutte_2024}. 

P2P energy trading and market clearance requires the coordination of the hubs in the network for the dispatch and operation of the various energy sources present in the different hubs. Centralized methods cast the trading in energy hub networks as a multi-objective optimization problem to balance specific costs, implement demand response program, etc.~\cite{Scala:2014,Yang:2016} and use advanced algorithms to tackle computation and scalability issues~\cite{Aghtaie:2014}. More commonly, the central optimization is solved in a distributed manner via strategies such as Lagrange relaxation~\cite{KRISHNAMOORTHY202172}, consensus ADMM~\cite{BEHRUNANI2024105922, Wang:2020} or as a distributed cooperative game~\cite{Fan:2018}. The P2P energy market can also be cast as a bi-level Stackelberg game~\cite{Wei:2017, Nasiri:2020}, or a non-cooperative game in which each prosumer has locally decoupled objectives and energy trading is incorporated through coupling reciprocity constraints~\cite{9143658, TUSHAR201910}. Designing prices that actively incentivize participation in the market by ensuring that the benefits of autonomous trading are equitably distributed is crucial since certain pricing choices may induce disproportionate costs for prosumers~\cite{le2020peer, BEHRUNANI20233751}. In~\cite{Fan:2018}, the benefits of P2P trading are distributed equally to all hubs, and in~\cite{Wang:2020} the transactive prices of each hub are derived based on internal operation for cost recovery. A sequential two-stage method is proposed in ~\cite{DARYAN:2022, BEHRUNANI20233751} wherein the optimal energy trades are computed in the first step, followed by the determination of bilateral trade prices between hubs in the second stage. Fairness among energy hubs in order to compute the trade prices and guarantees that agents always benefit by participating in the market is also considered in \cite{BEHRUNANI20233751}. Other pricing models include casting the problem as a Stackelberg game with the buyers as followers and sellers as the leaders~\cite{9535423}, or as a Markov decision process~\cite{8412581}. Several strategies consider product differentiation based on factors such as location~\cite{Sorin:2019} or individual preferences~\cite{8356100}.
 
A key challenge of P2P trading is its impact on the network. Trades between prosumers are carried out using the existing grid infrastructure and might jeopardize system reliability, for which network operators are responsible. A common way to address this issue is to include the network operator as another player that imposes the operational constraints of the network in the market~\cite{9732452,8957676}. Other approaches include decoupling the energy trades from the network constraints by requiring trade  approval by the system operator~\cite{8017536}, or using an opt-in flexibility market for prosumers that is ultimately cleared by the DSO~\cite{8572734}. In \cite{9744103}, the sensitivity of nodal voltage and network loss to power injections is used to evaluate the impact of P2P transactions on the network and the market is cleared in a distributed way. A sensitivity approach is also used in \cite{TARASHANDEH2024122240} to incorporate the network constraints in the P2P trading sub-problems without the network as a separate agent. Finally, \cite{9223743} employs generalized Nash bargaining theory to decompose the problem into two hierarchical subproblems, for social welfare maximization and for energy trading. 

Network operators also set grid usage tariffs for maintaining the grid and covering losses caused by P2P trades. In addition to fair trade prices, these tariffs are key to adequately compensate network losses, and prevent undesired congestion. A typical approach in literature is the application of Distributional Locational Marginal Prices to compute network usage fees~\cite{8938817}\cite{8957676}. In \cite{loss_allocation} and \cite{8630697}, the authors propose numerous alternative methods for designing exogenous network charges to reflect losses, network congestion, and utilization fees. The costs are assigning uniformly, based on electrical distance, and by geographical zones. Additionally, \cite{loss_allocation} also models interactions between the transmission and distribution grids. Electrical distance is determined by network topology using various methods. Most frequently, it is computed using the Power Transfer Distribution Factor (PTDF) which also used in \cite{9102274, ZARE2024122527} to account for power losses and network fees in P2P trading. The interaction between grid operators and prosumers is modelled as a Stackelberg game in \cite{9804862}, later converted into a single-level mixed integer quadratic program. The strategy in \cite{ZARE2024122527} also examines price distribution between buyers and sellers. 

While several studies focus on computing network tariffs and loss allocation in P2P markets, many rely on fixed policies like electrical or geographical distance. These methods often overlook factors such as prosumer type, total demand, or external conditions like weather. They also ignore the dynamic feedback between tariffs and trade values, and the sensitivity of trades to network tariffs. These strategies can unfairly penalize hubs with high grid usage costs and fail to account for their impact on operational costs, discouraging participation. Finally, while the computation of bilateral trading prices and network tariffs has been studied separately, their integration is largely unexplored. This work proposes a strategy to recover the cost of excess losses from P2P trading between hubs while minimizing hub costs. It accounts for network constraints but eliminates traditional loss allocation methods for tariff computation. Instead, it directly computes loss-aware network prices, leveraging tariff influence on trades to fairly distribute network costs and penalize inefficient, high-loss trades. We model this as a hierarchical Stackelberg game, combining network tariff design with optimal trade computation. Using distributed optimization, we propose a scalable, privacy-preserving distributed hypergradient-descent algorithm to solve the game. Additionally, we calculate fair bilateral P2P trading prices to redistribute benefits between hubs and incentivize market participation. By separating network and trade price computations, our approach reduces computational load while ensuring efficiency and fairness. Our contribution is two-fold:
\begin{enumerate}[(i)]

\item We formulate the problem of determining loss-aware network tariffs for P2P trading between energy hubs as a single-leader, multi-follower bilevel game. 

\item We design a scalable ADMM-based distributed hypergradient-descent algorithm to solve the bilevel game and compute network tariffs. Additionally, the framework includes a mechanism to compute fair bilateral trading prices while preserving the hierarchical structure, maintaining privacy, and reducing computational load.
\end{enumerate}
Finally, we illustrate and validate the proposed pricing mechanism through numerical simulations on a multi-hub network, using realistic models of electricity networks, energy hubs, and demand data under varying conditions. This paper is organized as follows: In Section II, the problem formulation and the model of the energy hubs, and the distribution network are presented. In Section III, the loss-aware network pricing strategy is elaborated and in Section IV, the computation of the trading pricing between hubs and the complete optimization framework is presented. Numerical case study and simulation results are presented in Section VI. Section VII concludes this paper and outlines directions for future research.

\vspace{-0.2cm}
\section{Problem formulation}
\begin{figure}
\centering
\includegraphics[width=89mm]{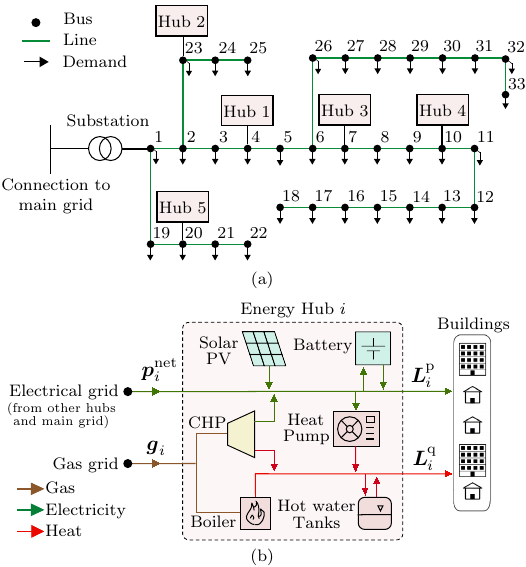}
\vspace{-0.5cm}
  \caption{(a) Illustration of the IEEE 33 bus benchmark test system with 5 energy hubs connected at different busses. (b) An example energy hub. }
  \label{fig:ch5_elect_grid_hub}
 \end{figure} 
\subsection{Energy hub modeling and optimization}
We consider a network of $H$ energy hubs labeled by $i \in \mathcal{H} := \{1,\dots,H\}$. The hubs are connected to the electricity and gas grid. Each energy hub is equipped with generation, conversion and storage devices that use energy from the grids to fulfill the aggregated electricity and heating load demands of the consumers connected to the hub. The demands are considered to be uncontrolled, and act as a measured disturbance for the hub controller. The hubs may contain renewable generation sources such as solar photovoltaic and solar thermal collectors, which generate electricity and/or heat, and conventional generation through combined Heat and Power (CHP) plants, which use gas to produce both electricity and heat. Hubs can also include energy converters such as gas boilers and heat pumps, that generate thermal energy using gas and electricity. Finally, the hubs may also comprise storage devices such as batteries to store electrical energy and water tanks to store thermal energy. The hubs can trade electrical energy with other hubs in the network via the electricity grid. We assume there is no global thermal grid or thermal energy trading among the hubs and that the devices in each hub are sized such that the thermal demand can be completely met locally at all times by conversion or storage. A thermal grid could be included, but it would introduce complex loss models and bilinearities. To keep the analysis simple, we exclude it from consideration. Fig. \ref{fig:ch5_elect_grid_hub}(a) shows the IEEE 33 bus benchmark test network with five energy hubs connected to it and \ref{fig:ch5_elect_grid_hub}(b) an example of an energy hub.

Consider a time horizon $\mathcal{T} = \{0,\dots,T-1\}$ and let $\mathcal{N}_i$ denote the set of devices of the energy hub $i$. The dynamics of each device $n \in \mathcal{N}_i$ are modeled as the following discrete-time linear state-space system that describes the evolution of its internal energy state $\boldsymbol{x}_{t,n}$:
\begin{equation}
\label{eq:ch5_hub_dynamics}
\left.\begin{array}{c}
\begin{aligned}
\boldsymbol{x}_{t+1,n}&=A_{n} \boldsymbol{x}_{t,n} + B_{n} \boldsymbol{u}_{t,n} + D_{n}  \boldsymbol{d}_{t,n}, \\
\boldsymbol{x}_{t,n} &\in \mathcal{X}_{n}, \ 
\boldsymbol{u}_{t,n} \in \mathcal{U}_{n},
\end{aligned}
\end{array}\right\} \forall t \in \mathcal{T},
\end{equation}
where $\boldsymbol{d}_{t,n}$ are the exogenous disturbances (e.g. solar radiation) acting on the system and $\boldsymbol{u}_{t,n}$ is the vector of control inputs for device $n$ at time $t$ (e.g. fuel absorbed, energy output, etc.). $A_{n},B_{n}$ and $D_{n}$ are the state, input and disturbance matrices, and $\mathcal{X}_{n}$ and $\mathcal{U}_{n}$ are the state and input constraint set, respectively. The vector $\boldsymbol{u}_{t,n}$ is defined as 
\begin{equation}
\nonumber
\begin{aligned}
\begin{array}{c c c }
\boldsymbol{u}_{t,n} = 
\left[ \begin{array}{c c c c c}
\boldsymbol{u}_{t,n}^{\mathrm{g,in}},& 
\boldsymbol{u}_{t,n}^{\mathrm{p,in}},&
\boldsymbol{u}_{t,n}^{\mathrm{q,in}},&
\boldsymbol{u}_{t,n}^{\mathrm{p,out}},&
\boldsymbol{u}_{t,n}^{\mathrm{q,out}}\end{array} \right]^{\text{T}},
\end{array}
\end{aligned}
\end{equation}
where $\boldsymbol{u}_{t,n}^{\mathrm{g,in}}$, $\boldsymbol{u}_{t,n}^{\mathrm{p,in}}$ and $\boldsymbol{u}_{t,n}^{\mathrm{q,in}}$ are the gas, electricity and heating input to the device at time $t$, respectively, and $\boldsymbol{u}_{t,n}^{\mathrm{p, out}}$ and $\boldsymbol{u}_{t,n}^{\mathrm{q,out}}$ are the electricity and heating outputs. For any variable $\boldsymbol{v}_{t\mathrm{,a}}$ at time $t$, the vector variable $\boldsymbol{v}_{\mathrm{a}}$ collects all the variables over the complete horizon $\mathcal{T}$ such that $\boldsymbol{v}_{\mathrm{a}}=\{\boldsymbol{v}_{t\mathrm{,a}}\}_{t \in \mathcal{T}}$. 

The energy hub's internal network is characterized by the electricity, heating and gas energy balance constraints. The electricity energy balance constraint for hub $i$ is given by
\begin{subequations}
\label{eq:ch5_hub_balance}
\begin{equation}
\label{eq:ch5_hub_elecbalance}
\begin{aligned}
\boldsymbol{L}^{\mathrm{p}}_{i} = \boldsymbol{e}^{\mathrm{out}}_{i} - \boldsymbol{e}^{\mathrm{in}}_{i} + \sum_{n \in \mathcal{N}_{i}} \left(\boldsymbol{u}_{n}^{\mathrm{p,out}} - \boldsymbol{u}_{n}^{\mathrm{p,in}}\right) + \sum_{j\in \mathcal{H}} \boldsymbol{p}_{ij}^{\mathrm{tr}},
\end{aligned}
\end{equation}
where $\boldsymbol{L}^{\mathrm{p}}_{i}$ is the total electricity demand of the consumers supplied by the energy hub $i$, $\boldsymbol{e}^{\mathrm{out}}_{i} \geq 0 $ and $\boldsymbol{e}^{\mathrm{in}}_{i}\geq 0$ is the electricity purchased from and sold to the electricity grid, and $\boldsymbol{p}_{ij}^{\mathrm{tr}}$ is the energy imported from hub $j$ to hub $i$. We define $\boldsymbol{p}^{\mathrm{tr}}_{ii} = 0, \forall i \in \mathcal{H}$. The total energy imported from all the hubs in the network is $\sum_{j \in \mathcal{H}}\boldsymbol{p}_{ij}^{\mathrm{tr}}$. Note that the energy balance also includes the electrical energy input and output from the hub devices. Similarly, the thermal energy balance constraint for hub $i$ is given by
\begin{equation}
\label{eq:ch5_hub_heatbalance}
\begin{aligned}
\boldsymbol{L}^{\mathrm{q}}_{i} &= \sum_{n \in \mathcal{N}_{i}} \left(\boldsymbol{u}_{n}^{\mathrm{q,out}} - \boldsymbol{u}_{n}^{\mathrm{q,in}}\right),
\end{aligned}
\end{equation}
where $\boldsymbol{L}^{\mathrm{q}}_{i}$ is the total thermal demand served by the energy hub $i$. The demand is fulfilled solely by the thermal input and output from the devices in the hub. We consider a simplified model of the thermal dynamics with no thermal loss within each hub's thermal network here\footnote{A comprehensive model that considers temperature constraints, pipe dynamics, etc. is left as a topic of future work.}. Finally, the total gas demand of the hub, $\boldsymbol{g}_{i}$, is given by
\begin{equation}
\label{eq:ch5_hub_gasbalance}
\begin{aligned}
\boldsymbol{g}_{i} = \sum_{n \in \mathcal{N}_{i}} \boldsymbol{u}_{n}^{\mathrm{g,in}}.
\end{aligned}
\end{equation}
\end{subequations}

Energy trades are subject to a reciprocity constraint which ensures that the energy traded from $i$ to $j$ is the inverse of that traded from $j$ to $i$,
\begin{equation}
\label{eq:ch5_hub_recip}
\boldsymbol{p}_{ij}^{\mathrm{tr}}  + \boldsymbol{p}_{\mathrm{ji}}^{\mathrm{tr}}  = 0, \ \ \ \forall j\in \mathcal{H}.
\end{equation}
Let $\boldsymbol{p}^{\mathrm{net}}_{i}$ be the net energy injected into the hub by the electrical grid at the bus that the hub is connected to. This is the sum of the total energy purchased from and sold to the electricity grid, and the energy traded with the other hubs,
\begin{equation}
\label{eq:ch5_hub_net}
\boldsymbol{p}^{\mathrm{net}}_{i} = \boldsymbol{e}^{\mathrm{out}}_{i} - \boldsymbol{e}^{\mathrm{in}}_{i} + \sum_{j\in \mathcal{H}} \boldsymbol{p}_{ij}^{\mathrm{tr}}.
\end{equation}
It impacts the network dynamics and the energy balance of the grid. Let $\boldsymbol{p}_{i} = \{ \left\{\boldsymbol{u}_{n}, \boldsymbol{x}_{n}\right\}_{\forall n \in \mathcal{N}_i},\ \boldsymbol{e}^{\mathrm{in}}_{i}, \ \boldsymbol{e}^{\mathrm{out}}_{i}, \ \boldsymbol{g}_{i}, \ \boldsymbol{p}^{\mathrm{net}}_{i}, \left\{\boldsymbol{p}^{\mathrm{tr}}_{ij} \right\}_{\forall j\in \mathcal{H}} \}$ collect all the variables for energy hub $i$, and $\mathscr{P}_{i}:= \left \{ \boldsymbol{p}_{i}~|~ \text{(\ref{eq:ch5_hub_dynamics}, \ref{eq:ch5_hub_balance}, \ref{eq:ch5_hub_recip}, \ref{eq:ch5_hub_net}) hold}\right \}$ its constraint set. 

For hub $i$, the cost of trading with any hub $j$ is given by $${\boldsymbol{c}^T_{ij}} \boldsymbol{p}_{ij}^{\mathrm{tr}} + {\boldsymbol{\gamma}^T_{ij}} {\lvert \boldsymbol{p}_{ij}^{\mathrm{tr}} \rvert},$$ where $\boldsymbol{c}_{ij}$ is the vector the bilateral trade prices with hub $j$ at each time $t\in \mathcal{T}$, and $\boldsymbol{\gamma}_{ij}$ is a trading tariff imposed by the network operator for the use of the grid infrastructure for P2P trading. 

The goal of the each energy hub optimization is to minimize the total operating costs while ensuring that the hub constraints are satisfied. The operating costs comprise the cost of the energy consumed from the electricity and gas grids, the cost of the energy fed into the electricity grid during periods of high production and the cost of bilateral trades with all the other hubs in the network. The resulting economic dispatch problem over the horizon $\mathcal{T}$ can be compactly written as:
\begin{align} 
    \min_{\boldsymbol{p}_i} & \ 
    \scalebox{0.93}{$\underbrace{
    {\boldsymbol{c}^{\mathrm{out}}_{\mathrm{e}}}^T \boldsymbol{e}^{\mathrm{out}}_{i} 
    - {\boldsymbol{c}^{\mathrm{in}}_{\mathrm{e}}}^T\boldsymbol{e}^{\mathrm{in}}_{i} 
    + {\boldsymbol{c}^{\mathrm{out}}_{\mathrm{g}}}^T\boldsymbol{g}_{i} 
    + \sum_{j \in \mathcal{H}} 
    \left({\boldsymbol{c}^T_{ij}} \boldsymbol{p}^{\mathrm{tr}}_{ij} 
    + {\boldsymbol{\gamma}^T_{ij}} \lvert \boldsymbol{p}_{ij}^{\mathrm{tr}} \rvert 
    \right)}_{\boldsymbol{J}_{i}(\boldsymbol{p}_{i},\boldsymbol{c}_{i})}$} \notag \\
    \text{s.t.} & \quad \boldsymbol{p}_i \in \mathscr{P}_i,\label{eq:ch5_energyhub_optimization}
\end{align}
where $\boldsymbol{c}^{\mathrm{out}}_{\mathrm{e}}$, and $\boldsymbol{c}^{\mathrm{out}}_{\mathrm{g}}$ is the per unit price vector of electricity and gas consumed from the grid, respectively, and $\boldsymbol{c}^{\mathrm{in}}_{\mathrm{e}}$, is the feed-in price for electricity paid to the hub for feeding excess electricity into the grid. We assume dynamic prices for electricity and gas that vary at each time $t$, and is known by the hub controller for the complete horizon $\mathcal{T}$. In practice, electricity prices may vary across hubs to reflect locational marginal pricing or different contracts with grid operators. For simplicity, we assume uniform prices for all hubs in the network.
\vspace{-0.3cm}
\subsection{Distribution Network Optimization}

Consider the electrical distribution network, to which the energy hubs and consumers are physically connected as illustrated in Fig. \ref{fig:ch5_elect_grid_hub} (a). We assume that the network is a connected undirected graph $\mathcal{G} = (\mathcal{B},\mathcal{L})$ comprising $B$ busses, indexed by $b \in \mathcal{B} := \{1, 2,..., B\}$, connected pairwise by $L$ power lines, indexed by $l \in \mathcal{L} := \{1, 2,..., L\}$, each connected to two busses, respectively. We assume that hub in the network is connected to a single bus but the same bus may supply more than one hub. We consider a linear approximation of power-flow equations, which is standard in the literature of P2P markets\cite{loss_allocation, 9732452}.  

The active power balance for bus $b \in \mathcal{B}$ is given by
\begin{equation}
\label{eq:ch5_net_nodeactive}
\boldsymbol{p}_b^{\mathrm{mg}}=\sum_{l \in \mathcal{L}_b} \left(\boldsymbol{f}_{l}^{\mathrm{p}} +0.5 \cdot \boldsymbol{w}_{l} \right) + \sum_{i \in \mathcal{H}_b} \boldsymbol{p}^{\mathrm{net}}_{i} + \boldsymbol{F}^{\mathrm{p}}_{b} 
\end{equation}
where $\mathcal{L}_b \subseteq \mathcal{L}$ and $\mathcal{H}_b \subseteq \mathcal{H}$ denote the set of lines and the set of energy hubs connected to the bus $b \in \mathcal{B}$, respectively, $\boldsymbol{p}_b^{\mathrm{mg}}$ the active power exchanged between bus $b$ and the main grid, $\boldsymbol{F}_b^{\mathrm{p}}$ the active power demand of the consumers connected to the bus $b$ other than energy hubs, $\sum_{i \in \mathcal{H}_b} \boldsymbol{p}^{\mathrm{net}}_{i}$ the aggregated power injected into the hubs connected to the bus $b$, and $\boldsymbol{f}_{l}^{p}$ and $\boldsymbol{w}_{l}$ the active power flow and the energy loss in the line $l \in \mathcal{L}$. The value of $\boldsymbol{f}_{l}^{p}$ is positive for a line $l = (b, c) \in \mathcal{L}$ if the power flows form $b$ to $c$ and negative otherwise. The line loss is transformed into nodal losses and incorporated into the active power balance by allocating half of the loss to each of the two busses that the line connects. The constraint imposed on the voltage and phase angle of each $b \in \mathcal{B}$ is 
\begin{equation}
\begin{aligned}
\label{eq:ch5_net_nodelimit}
\underline{\boldsymbol{\theta}}  \leq \boldsymbol{\theta}_b \leq \bar{\boldsymbol{\theta}} , \ \ \ \underline{\boldsymbol{v}} \leq \boldsymbol{v}_b \leq \bar{\boldsymbol{v}} ,
\end{aligned}
\end{equation}
where $\boldsymbol{v}_{b}$ and $\boldsymbol{\theta}_{b}$ are the voltage magnitude and the phase angle for bus $b$, and $\underline{\boldsymbol{v}}/\bar{\boldsymbol{v}}$ and $\underline{\boldsymbol{\theta}}/\bar{\boldsymbol{\theta}}$ be their lower/upper limits. We assume that a subset $\mathcal{B}_{\mathrm{mg}}$ of the set of busses are connected to the main grid and set, $\boldsymbol{p}_b^{\mathrm{mg}}=0, \ \forall \boldsymbol{b} \notin \mathcal{B}_{\mathrm{mg}}$.

The linearized power flow equations for each line $l = (b,c) \in \mathcal{L}$ from the perspective of bus $b$, are 
\begin{equation}
\label{eq:ch5_net_linelimit}
\begin{aligned}
& \boldsymbol{f}_{l}^{p} =B_{l}\left(\boldsymbol{\theta}_b-\boldsymbol{\theta}_c\right)-G_{l}\left(\boldsymbol{v}_b-\boldsymbol{v}_c\right),  \\
& \boldsymbol{f}_{l}^{q}=G_{l}\left(\boldsymbol{\theta}_b-\boldsymbol{\theta}_c\right)+B_{l}\left(\boldsymbol{v}_b-\boldsymbol{v}_c\right), \\
&\left(\boldsymbol{f}_{l}^{p}\right)^2+\left(\boldsymbol{f}_{l}^{q}\right)^2 \leq \bar{\boldsymbol{f}}_{l}^2
\end{aligned}
\end{equation}
where $B_{l}$ and $G_{l}$ denote the susceptance and conductance, respectively, of line $l$. Additionally, we impose conic line capacity constraints on each line, $\bar{\boldsymbol{f}}_{l}$. Finally, following \cite{loss_allocation, 8782597}, we employ a linear approximation the losses, $\boldsymbol{w}_l$, of line $l$, 
\begin{equation}
\label{eq:ch5_net_lineloss}
\boldsymbol{w}_l=\boldsymbol{M}_l\left|\boldsymbol{f}_{l}^{p}\right|+\boldsymbol{Q}_l.
\end{equation}
The coefficients of the linear approximation, $\boldsymbol{M}_l$ and $\boldsymbol{Q}_l$, can be estimated using least squares or linearizing around a certain range of flows~\cite{8782597}. Formulating the loss function in this way makes it possible to calculate losses without considering the direction of the flow. Let $\boldsymbol{\Gamma} =\{ \left\{\boldsymbol{p}_b^{\mathrm{mg}}, \boldsymbol{v}_b, \boldsymbol{\theta}_b \right\}_{\forall b \in \mathcal{B}},  \left\{\boldsymbol{f}_{l}^{p} , \boldsymbol{f}_{l}^{q} , \boldsymbol{w}_{l} \right\}_{\forall l \in \mathcal{L}}\}$ be the set of all variables associated with network, and $\mathscr{G}:=\{\boldsymbol{\Gamma}   ~|~(\ref{eq:ch5_net_nodeactive},\ref{eq:ch5_net_nodelimit},\ref{eq:ch5_net_linelimit},\ref{eq:ch5_net_lineloss}) \text{ hold}\}$ is the full constraint set of the network.

The goal of the distribution network is to minimize the total energy cost of imports from the main grid while ensuring that all network constraints are satisfied. The resulting network optimization problem over the horizon $\mathcal{T}$ can be compactly written as:
\begin{equation}%
\label{eq:ch5_network_optimization}
\min_{\boldsymbol{\Gamma}} \  \sum_{b \in \mathcal{B}} {\boldsymbol{c}^{T}_{\mathrm{mg}}} \boldsymbol{p}_b^{\mathrm{mg}} \ \ \text { s.t.} \ \ \boldsymbol{\Gamma} \in \mathscr{G}
\end{equation}
where $\boldsymbol{c}_{\mathrm{mg}}$ is the per unit price vector of importing energy from the main grid which can vary at each time $t$, and is known by the network operator for the complete horizon $\mathcal{T}$.

\vspace{-0.2cm}
\subsection{Market Setup}
The distribution market operator objective is to design the network price in such a way that the tariff imposed on the hubs for using the electricity grid infrastructure for P2P energy trading is sufficient to recover the cost of the additional line losses generated by the trades. The design mechanism also aims to ensure that the trade prices are fair, and incentivize trades that reduce losses and hub costs. Let $\boldsymbol{p}$ collect the optimization variables of the hubs $\boldsymbol{p} = \{\boldsymbol{p}_i\}_{i \in \mathcal{H}}$, and $\boldsymbol{\gamma}$ collect all the trading tariffs $\boldsymbol{\gamma}= \{ \boldsymbol{\gamma}_{ij}\}_{(i,j) \in \mathcal{H} \times \mathcal{H} }$. The network tariff for all trades is set by solving the following:
\begin{subequations}
\label{eq:ch5_bilevel}
\allowdisplaybreaks
\begin{align}
\label{eq:ch5_bilevel_obj}  \min_{\boldsymbol{\gamma}} &  \underbrace{ \sum_{(i,j) \in \mathcal{H} \times \mathcal{H} } \boldsymbol{\gamma}_{ij}^{T}\lvert \boldsymbol{p}_{ij}^{\mathrm{tr}} \rvert +  \boldsymbol{\gamma}_{ij}^2}_{J\left(\boldsymbol{\gamma},\boldsymbol{p}\right)} \\
\label{eq:ch5_bilevel_con1}\text { s.t. }  &  \sum_{(i,j) \in \mathcal{H} \times \mathcal{H} } \boldsymbol{\gamma}_{ij}^{T} \lvert \boldsymbol{p}_{ij}^{\mathrm{tr}} \rvert \geq {\boldsymbol{c}^{\mathrm{out}}_{\mathrm{e}}}^{T} \left(\sum_{l \in \mathcal{L}} \boldsymbol{w}_{l} - \sum_{l \in \mathcal{L}} \boldsymbol{w}^{\mathrm{NT}}_{l}\right)\\
\label{eq:ch5_bilevel_con2}&\boldsymbol{\gamma}_{ij} \geq 0\\
\label{eq:ch5_bilevel_con3}&\boldsymbol{\gamma}_{ij} = \boldsymbol{\gamma}_{ji}, \ \ \  \forall (i,j) \in \mathcal{H} \times \mathcal{H} 
\end{align}
where $\boldsymbol{w}^{\mathrm{NT}}_{l}$ are the nominal line losses that would have occurred in the network with no P2P trading. The first constraint ensures that the total amount collected from the trading tariff recovers any additional costs of losses due to P2P trading. The second constraint ensures positivity of the tariffs to maintain convexity of both this optimization as well as the hub optimization \eqref{eq:ch5_energyhub_optimization}. The final constraint ensures both hubs in the trade are penalized equally for the losses. The objective function of the grid operator minimizes the total cost to the hubs and includes a regularization term to minimize the value of $\gamma$ as well as to make the problem strongly convex. Let $\mathscr{M} :=\{\boldsymbol{\gamma} ~|~(\ref{eq:ch5_bilevel_con1}), (\ref{eq:ch5_bilevel_con2}), (\ref{eq:ch5_bilevel_con3}) \text{ hold}\}$ be the constraint set of the market setup. 
\end{subequations}

\vspace{-0.3cm}
\section{Loss aware Network pricing}

The optimization of the grid operator is coupled to that of the energy hubs by the term $\sum_{(i,j) \in \mathcal{H} \times \mathcal{H} } \boldsymbol{\gamma}_{ij}^T\lvert \boldsymbol{p}_{ij}^{\mathrm{tr}} \rvert$ leading to a single-leader multi-follower bilevel game. The market operator acts as the game leader that sets the trading tariff, and the hubs and the distribution network are the followers that compute the optimal hub and network setpoints in response. To compute an equilibrium of this game, we propose a distributed projected hypergradient descent algorithm adapted from \cite{big_hype} to fit our problem formulation.

The proposed scheme is summarized in Alg. \ref{alg1:ADMM_hypergrad} and consists of two nested loops. In the inner loop, the followers(the energy hubs and the network) receive the current grid tariff and compute their optimal solution as well as their sensitivity to the leader's variables in a distributed manner using Alg. \ref{alg:admm_sens}. In the outer loop, the leader(market operator) uses these to perform a projected hyper-gradient step resulting in a new tariff; this is communicated to the followers and the process is repeated. This iteration continues until a termination criterion is satisfied. The results in \cite{big_hype} then imply that Alg. \ref{alg1:ADMM_hypergrad} is guaranteed to converge for appropriately chosen step sizes, $\left\{\alpha^k, \beta^k\right\}_{k \in \mathbb{N}}$, and tolerance, $\sigma$. It states that $\left\{\alpha^k\right\}_{k \in \mathbb{N}}$ must be nonnegative, non-summable and square-summable, $\{\beta^k\}_{k \in \mathbb{N}}=1$, and $\sum_{k=0}^{\infty} \alpha^k \sigma<\infty$.


The computation of fair bilateral trading prices $\boldsymbol{c} = \{\boldsymbol{c}_{ij}\}_{\forall (i,j) \in \mathcal{H} \times \mathcal{H}} $ has been discussed in \cite{BEHRUNANI20233751} wherein the authors prove that the trading game of this form between hubs has a unique equilibrium solution which does not depend on the choice of the bilateral trading prices. Therefore,
the trading price $\boldsymbol{c}$ will have no impact on the optimal hub and network strategy. Here, we set $\boldsymbol{c}$ to 0 to compute the optimal economic dispatch of the hubs and the network, then determine the trading prices as summarized in Section IV.  

\begin{figure}[t] 
\label{alg1:ADMM_hypergrad}
\flushleft
\hrule
\smallskip
\textsc{Algorithm III}: ADMM-based Hyper-gradient Descent
\smallskip
\hrule
\smallskip
\textbf{Parameters: } $\overline{\boldsymbol{k}}$, $\overline{\boldsymbol{w}}$, step sizes $\left\{\boldsymbol{\alpha}^k, \boldsymbol{\beta}^k\right\}_{k \in \mathbb{N}}$, tolerances $\boldsymbol{\epsilon}, \boldsymbol{\sigma}$\\ [.2em]
\textbf{Initialization: } $\boldsymbol{k}=0$  , $\boldsymbol{c} = 0$, Tariff $\boldsymbol{\gamma}^{0}$ \\
\hspace*{6.4em} Trades $\boldsymbol{p}^{\mathrm{net},0}, [\boldsymbol{p}^{\mathrm{tr},0}_{ij}]_{\forall(i,j) \in \mathcal{H} \times \mathcal{H} }$ \\
\hspace*{6.4em} Duals $\boldsymbol{\mu}^{0}, [\boldsymbol{\lambda}^{0}_{ij}]_{\forall(i,j) \in \mathcal{H} \times \mathcal{H} }$ \\
 [.2em]
\textbf{Iterate until convergence:}  \\[.2em]
$
\left\lfloor
\begin{array}{l}
\text{1. Market operator hyper-gradient step :}\\\hspace*{0.1em}
\left\lfloor
\begin{array}{l}
\text{a. Compute the hyper-gradient using \eqref{eq:ch5_hypergradient}:}\\[0.2em]
\nabla J\left(\boldsymbol{\gamma}, \boldsymbol{p}^{\star}\right) =\nabla_{\boldsymbol{\gamma}}J\left(\boldsymbol{\gamma},\boldsymbol{p}^{\star}\right)+ \boldsymbol{s}_{\boldsymbol{\gamma}}(\boldsymbol{p}^{\star})^{T} \nabla_{\boldsymbol{p}}J\left(\boldsymbol{\gamma},\boldsymbol{p}^{\star}\right) \\ [0.3em]
\text{b. Perform projected hyper-gradient descent step (\ref{eq:ch5_projection}):}\\
\qquad \begin{array}{r l}\widehat{\boldsymbol{\gamma}}  &= \boldsymbol{\gamma}^k-\boldsymbol{\alpha}^k \nabla J^k\\
\boldsymbol{\gamma}^{k+1} &=\boldsymbol{\gamma}^k+\boldsymbol{\beta}^k\left(\mathbb{P}_{\mathscr{M}}\left[\widehat{\boldsymbol{\gamma}}\right]-\boldsymbol{\gamma}^k\right)
\end{array}\\
\end{array}
\right.
\hspace*{1em}\\[3em]
\text{2. Dist. network opt. and sensitivity comp. using \ref{alg:admm_sens}}\\ \hspace*{0.1em}
\left\lfloor
\begin{array}{l}
\text{\textit{Parameters}: $\overline{\boldsymbol{w}}$, $\boldsymbol{\epsilon}$}\\ 
\text{\textit{Inputs}: Trades $\boldsymbol{p}^{\mathrm{tr},k}_{ij}, \boldsymbol{p}^{\mathrm{net},k}$, Duals $\boldsymbol{\lambda}^{k}_{ij},\boldsymbol{\mu}^{k}$, Tariff $\boldsymbol{\gamma}^{k+1}$}\\ 
\text{\textit{Outputs}: Trades $ \boldsymbol{p}^{\mathrm{tr},k+1}_{ij}$, $ \boldsymbol{p}^{\mathrm{net},k+1}$ Duals $\boldsymbol{\lambda}^{k+1}_{ij}$,$\boldsymbol{\mu}^{k+1}$  }\\
\hspace*{3.8em} \text{Sensitivities $\boldsymbol{s}_{\boldsymbol{\gamma}}(\boldsymbol{p}_i^{k+1}) \ \forall i \in \mathcal{H}$}
\\
\hspace*{3.8em} \text{Optimal setpoints $\boldsymbol{p}^{k+1}$ and $\boldsymbol{\Gamma}^{k+1}$}\\
\end{array}
\right.
\hspace*{1em}\\[3em]
\end{array}
\right. 
$\\[-0.6em]
\hspace*{0.8em} $\boldsymbol{k} \leftarrow \boldsymbol{k}+1$\\[.2em]
\textbf{Return: } Optimal setpoints $\boldsymbol{p}^*$, $\boldsymbol{\Gamma}^*$, tariff $\boldsymbol{\gamma}^*$, duals $\boldsymbol{\lambda}^*_{ij}$,$\boldsymbol{\mu}^*$
\smallskip
\hrule
\end{figure}

\vspace{-0.3cm}
\subsection{Leader optimization through hyper-gradient}
We approach the market operator's optimization problem \eqref{eq:ch5_bilevel} by substituting the solution of the followers optimization, $\boldsymbol{p}^{\star}$ and $\boldsymbol{\Gamma}^{\star}$ in \eqref{eq:ch5_bilevel_obj} and minimizing the resulting objective, $J\left(\boldsymbol{\gamma},\boldsymbol{p}^{\star}\right)$. By projected hyper-gradient descent, the hyper-gradient is given by:
\begin{equation}
\label{eq:ch5_hypergradient}
\begin{aligned}
\nabla J\left(\boldsymbol{\gamma}, \boldsymbol{p}^{\star}\right) & =\nabla_{\boldsymbol{\gamma}}J\left(\boldsymbol{\gamma},\boldsymbol{p}^{\star}\right)+ \boldsymbol{s}_{\boldsymbol{\gamma}}(\boldsymbol{p}^{\star})^{\top} \cdot \nabla_{\boldsymbol{p}}J\left(\boldsymbol{\gamma},\boldsymbol{p}^{\star}\right) \\
& = 4 \boldsymbol{\gamma}  +\lvert \boldsymbol{p}^{\star} \rvert + \boldsymbol{s}_{\boldsymbol{\gamma}} (\boldsymbol{p}^{\star})^{\top} \cdot \boldsymbol{\gamma} \cdot \text{sgn}(\boldsymbol{p}^{\star})
\end{aligned}
\end{equation}
where $\boldsymbol{s}_\gamma(p^{\star})^{\top}$ is the Jacobian of the solution $\boldsymbol{p}^{\star}$ with respect to $\boldsymbol{\gamma}$ computed in the inner loop (step 2.) together with the optimal solution. 
In (b.), the hypergradient is used to perform a gradient step with step size $\alpha_k$ and projected into the constraint set using the projection operator which computes the least square solution. The projection operator is written as: 
\begin{equation}
\label{eq:ch5_projection} \mathbb{P}_{\mathscr{M}}\left[\widehat{\boldsymbol{\gamma}}\right] = \ \arg \min_{\boldsymbol{g}} \left\| \boldsymbol{g}-\widehat{\boldsymbol{\gamma}}\right\|_2^2 \ \ \text { s.t.} \ \ \boldsymbol{g} \in \mathscr{M}
\end{equation}

The algorithm terminates if either the total payment collected from the hubs exceeds the cost of additional losses and changes by less than a tolerance $\sigma$ or if the maximum number of iterations $\bar{k}$ is reached. Since the algorithm is guaranteed to converge, the iteration limit is set only to prevent excessive computation time. If this limit is reached before convergence, we use either a predetermined constant tariff or the solution from the previous time step.

\begin{equation}
\nonumber
\scalebox{0.93}{$
\begin{aligned}
\left(\begin{array}{cc}
\begin{aligned}
     \left| \sum_{(i,j) \in \mathcal{H} \times \mathcal{H} } \left(\boldsymbol{\gamma}_{ij}^{kT}\lvert p_{ij}^{\text{tr,k}} \rvert - \boldsymbol{\gamma}_{ij}^{kT} \lvert p_{ij}^{\text{tr,k-1}} \rvert \right) \right| \leq \sigma \\
 c^{\mathrm{out}}_{\text{e}} \sum_{l \in \mathcal{L}} \left(w^{k}_{\text{l}} - w^{\text{NT}}_{\text{l}}\right) \leq  \sum_{(i,j) \in \mathcal{H} \times \mathcal{H} } \boldsymbol{\gamma}_{ij}^{kT} \lvert p_{ij}^{\text{tr,k}} \rvert
 \end{aligned}
\end{array} \right)
\text{\textbf{or }} k &\geq \bar{k}
\end{aligned}
$}
\end{equation}

At convergence, the tariffs and the optimal set points are returned and can be applied to the energy hubs and the network at each time step over the complete horizon $\mathcal{T}$. The payments can then be settled between the hubs and the grid operator for the trades. Additionally, the algorithm also returns the optimal duals. These can be used along with the optimal trade values and tariffs to warm start nest iteration to speed up the convergence.

\vspace{-0.2cm}
\subsection{Followers optimization and sensitivity computation}

The inner loop is the P2P market clearing problem that involves solving the optimization for each hub \eqref{eq:ch5_energyhub_optimization}, and of the distribution network \eqref{eq:ch5_network_optimization}. These problems are coupled through the shared transfer variables, $\boldsymbol{p}_{ij}^{\mathrm{tr}}$ and the active power injection $\boldsymbol{p}^{\mathrm{net}}_{i}$. Motivated by our earlier work \cite{BEHRUNANI2024105922}, we propose a distributed algorithm for solving this problem based on the Alternating Direction Method of Multipliers (ADMM).

ADMM creates local copies of each coupling variable for each agent, and a consensus variable communicated between them. For the P2P trade between hub $i$ and $j$, $\boldsymbol{p}_{ij}^{\mathrm{tr}}$, let $\boldsymbol{p}_{ij,i}^{\mathrm{tr}}$ and $\boldsymbol{p}_{ij,j}^{\mathrm{tr}}$ be the local copies for Hub $i$ and Hub $j$, respectively, and $\boldsymbol{z}_{ij}^{\mathrm{tr}}$ be the consensus variable. Similarly, for the power injection of hub $i$ into the network, $\boldsymbol{p}_{i}^{\mathrm{net}}$, let $\boldsymbol{p}_{i,i}^{\mathrm{net}}$ and $\boldsymbol{p}_{i,n}^{\mathrm{net}}$ be the local copies for hub $i$ and the network, respectively, and $\boldsymbol{z}_{i}^{\mathrm{net}}$ be the consensus variable. Equality constraint of the form $\boldsymbol{p}_{ij,i}^{\mathrm{tr}} - \boldsymbol{z}_{ij}^{\mathrm{tr}} = 0$ is added for each local copy to align the value to the consensus variable. The resulting optimization for hub $i$ formulated using the augmented Lagrangian is:
\begin{align}
&\nonumber\min_{\boldsymbol{p}_i} \
\scalebox{0.92}{${\boldsymbol{c}^{\mathrm{out}}_{\mathrm{e}}}^T \boldsymbol{e}^{\mathrm{out}}_{i} - {\boldsymbol{c}^{\mathrm{in}}_{\mathrm{e}}}^T\boldsymbol{e}^{\mathrm{in}}_{i} + {\boldsymbol{c}^{\mathrm{out}}_{\mathrm{g}}}^T\boldsymbol{g}_{i} + $}\sum_{j \in \mathcal{H}} \scalebox{0.92}{$\Big({\boldsymbol{c}^T_{ij}} \boldsymbol{p}^{\mathrm{tr}}_{ij} + {\boldsymbol{\gamma}^T_{ij}} \lvert \boldsymbol{p}_{ij}^{\mathrm{tr}} \rvert $}
\\ \nonumber& \ \ \ \ \ \scalebox{0.92}{$+
\boldsymbol{\lambda}_{ij,i}^{\mathrm{T}}\left(\boldsymbol{p}^{\mathrm{tr}}_{ij,i}-\boldsymbol{z}^{\mathrm{tr}}_{ij}\right) + \frac{\boldsymbol{\rho}}{2}\left\|\boldsymbol{p}^{\mathrm{tr}}_{ij,i}-\boldsymbol{z}^{\mathrm{tr}}_{ij}\right\|_2^2 + \boldsymbol{\lambda}_{ji,i}^{\mathrm{T}}\left(\boldsymbol{p}^{\mathrm{tr}}_{ji,i}-\boldsymbol{z}^{\mathrm{tr}}_{ji}\right) $}
\\ \nonumber& \ \ \ \ \ 
 \scalebox{0.92}{$+ \frac{\boldsymbol{\rho}}{2}\left\|\boldsymbol{p}^{\mathrm{tr}}_{ji,i}-\boldsymbol{z}^{\mathrm{tr}}_{ji}\right\|_2^2 \Big) +
\boldsymbol{\mu}_{i,i}^{\mathrm{T}}\left(\boldsymbol{p}^{\mathrm{net}}_{i,i}-\boldsymbol{z}^{\mathrm{net}}_{i}\right) + \frac{\boldsymbol{\rho}}{2}\left\|\boldsymbol{p}^{\mathrm{net}}_{i,i}-\boldsymbol{z}^{\mathrm{net}}_{i}\right\|_2^2$}
\\
\label{eq:ch5_hub_optimization_admm}&\text{s.t.} \ \  \boldsymbol{p}_i \in \mathscr{P}_i,
\end{align}
where $\lambda_{ij,i}$ and $\mu_{i,i}$ are the Lagrange dual variables, and $\rho_{\geq 0}$ is the augmented Lagrangian penalty parameter. Analogously, the optimization for the network is written as:
\begin{align}
\nonumber \min_{\boldsymbol{\Gamma}}  &  \sum_{b \in \mathcal{B}} \scalebox{0.92}{$\boldsymbol{c}^T_{\mathrm{mg}} \boldsymbol{p}_b^{\mathrm{mg}}$} +
 \sum_{i \in \mathcal{H}} \scalebox{0.92}{$\left( \boldsymbol{\mu}^{\mathrm{T}}_{i,n}\left(\boldsymbol{p}^{\mathrm{net}}_{i,n}-\boldsymbol{z}^{\mathrm{net}}_{i}\right) + \frac{\boldsymbol{\rho}}{2}\left\|\boldsymbol{p}^{\mathrm{net}}_{i,n}-\boldsymbol{z}^{\mathrm{net}}_{i}\right\|_2^2 \right)$} \\
\label{eq:ch5_network_optimization_admm}\text { s.t. } &  \boldsymbol{\Gamma} \in \mathscr{G},
\end{align} 
where $\boldsymbol{\mu}_{i,n}$ is the Lagrange dual variable. 


The Consensus ADMM algorithm is summarized in Alg. \ref{alg:admm_sens}. The algorithm is initialized with inputs from the outer loop for the trading tariff, dual variables, and consensus variables. Each hub and the network operator independently solve their optimization problems to compute the optimal dispatch and trade. Next, the local copies are communicated between agents, and the consensus and dual variables are updated. The optimization problems are then resolved with the updated values and the process repeats; this cycle continues until all shared trade variables converge. The algorithm terminates once the primal residuals for all the hubs and the network are lower than the tolerance $\epsilon$ or if the number of iterations, $w$, reaches a max $\bar{\boldsymbol{w}}$. This is written as: 
\begin{equation}
\nonumber
 \left( \left\|\boldsymbol{r}^{\mathrm{prim}}_{n}\right\|^{2}_{2}, \ \left\|\boldsymbol{r}^{\mathrm{prim}}_{i}\right\|^{2}_{2} \leq \boldsymbol{\epsilon} \
 \forall i \in \mathcal{H}\right) \text{ or } \boldsymbol{w} \geq \bar{\boldsymbol{w}} ,
\end{equation}
where $\boldsymbol{r}^{\mathrm{prim}}_{i}$ and $\boldsymbol{r}^{\mathrm{prim}}_{\mathrm{n}}$ is are the primal residuals for hub $i$ and the network, respectively, given by
\begin{align}
   \nonumber\boldsymbol{r}^{\mathrm{prim}}_{i} &= \left[\left(\boldsymbol{p}^{\mathrm{tr}}_{ij,i} - \boldsymbol{z}^{\mathrm{tr}}_{ij} \right) \ \forall j \in \mathcal{H} , \left(\boldsymbol{p}^{\mathrm{net}}_{i,i} - \boldsymbol{z}^{\mathrm{net}}_{i} \right) \right]^T , \\
   \nonumber\boldsymbol{r}^{\mathrm{prim}}_{\mathrm{n}} &= \left[\left(\boldsymbol{p}^{\mathrm{net}}_{i,n} - \boldsymbol{z}^{\mathrm{net}}_{i} \right) \ \forall i \in \mathcal{H} \right]^T .
\end{align}
The iteration limit is set only to prevents excessive computation. If reached before convergence, hubs act on their local copies of the coupling variable, with the grid compensating for electrical mismatches. Following this, each hub independently computes the sensitivity of its solution with respect to the tariff, $\boldsymbol{s}_\gamma(p^{\star})^{\top}$. The desired Jacobian is the value of $\mathrm{\bold{d}}\boldsymbol{p}_i$ obtained by solving the following set of equations~\cite{amos2017optnet}:
.
\begin{equation}
\label{eq:ch5_final_sens}
\scalebox{0.9}{
$\begin{aligned}
\left[\begin{array}{>{\centering\arraybackslash$} p{1.2cm} <{$}  >{\centering\arraybackslash$} p{1.9cm} <{$} >{\centering\arraybackslash$} p{0.4cm} <{$}}
			\boldsymbol{\rho} & G_i^T & A_i^T \\
			\hspace{-0.4em}D\left(\boldsymbol{\lambda}_i^{\star}\right)G_i & D\left(G_i \boldsymbol{p}_i^{\star}-\boldsymbol{h}_i\right) & 0 \\
			A_i & 0 & 0
		\end{array}\right] \hspace{-0.4em} \left[\begin{array}{>{\centering\arraybackslash$} p{0.4cm} <{$}}
			\mathrm{\bold{d}}\boldsymbol{p}_i \\
			\mathrm{\bold{d}}\boldsymbol{\lambda} \\
			\mathrm{\bold{d}}\boldsymbol{\nu}
		\end{array}\right] \hspace{-0.3em} =
		- \hspace{-0.3em} \left[\begin{array}{>{\centering\arraybackslash$} p{0.8cm} <{$}}
			\hspace{-0.4em}\text{sgn}(\boldsymbol{p}_i^{\star}) \\
			0 \\
			0
\end{array}\right],
\end{aligned}$
}
\end{equation}
where $G$, $A$, and $h$ are derived by expressing the constraint set $\mathscr{P}_i$ in the general matrix form: $\mathscr{P}_i := \{ \boldsymbol{p}_i~|~G_i \boldsymbol{p}_i \leq \boldsymbol{h}_i, \ A_i \boldsymbol{p}_i = \boldsymbol{b}_i \}$. The algorithm returns the optimal set points for the hubs and network, the trade values, the hubs' sensitivity, and the optimal dual variables. The optimal trade and dual values also serve as inputs to the ADMM algorithm for a warm start in the next iteration of the outer hyper-gradient algorithm.

\begin{figure}[t] \label{alg:admm_sens}
\flushleft
\hrule
\smallskip
\textsc{Algorithm III-B}: Distributed Consensus ADMM
\smallskip
\hrule
\smallskip
\textbf{Parameters: } $\overline{\boldsymbol{w}}$, tolerance $\boldsymbol{\epsilon}$\\ [.2em]
\textbf{Initialize using inputs:} $\boldsymbol{w}=0$, $\boldsymbol{c} = 0$ $\boldsymbol{\gamma} \leftarrow$ $ \boldsymbol{\gamma}^{k+1} $\\
\hspace*{6em}Trades $\boldsymbol{z}^{\mathrm{net}} \leftarrow \boldsymbol{p}^{\mathrm{net},k}$,$  [\boldsymbol{z}^{\mathrm{tr}}_{ij} \leftarrow \boldsymbol{p}^{\mathrm{tr},k}_{ij}]_{\forall(i,j) \in \mathcal{H} \times  \mathcal{H}} $ \\
\hspace*{6em}Duals $\boldsymbol{\mu}^{0} \leftarrow \boldsymbol{\mu}^{k}$, $[\boldsymbol{\lambda}^{0}_{ij} \leftarrow \boldsymbol{\lambda}^{k}_{ij}]_{\forall(i,j) \in \mathcal{H} \times \mathcal{H} } $ \\ [.2em]
\textbf{Iterate until convergence:}  
\\[.2em]
\hspace*{.5em}$
\left\lfloor
\begin{array}{l}
\text{1. Solve \eqref{eq:ch5_hub_optimization_admm}) $\forall i \in \mathcal{H}$  and \eqref{eq:ch5_network_optimization_admm} for the network}\\ [.2em]
\text{2. Communicate locally computed values:}\\ \hspace*{0.5em}
\left\lfloor
\begin{array}{l}
\text{- $\boldsymbol{p}^{\mathrm{tr}}_{ij,i/j}$ , $\boldsymbol{p}^{\mathrm{tr}}_{ji,i/j}$ between hub $i$ and $j$ {\footnotesize $\forall (i,j) \in \mathcal{H} \times \mathcal{H} $}} \\ 
\text{- $\boldsymbol{p}^{\mathrm{net}}_{i,i/n}$ between hub $i$ and network {\footnotesize $\forall i \in \mathcal{H}$}}
\end{array}
\right. \\ [1em]
\text{3. Update all trades } \boldsymbol{z}^{\mathrm{tr}}_{ij} \text{ and } \boldsymbol{z}^{\mathrm{net}}_{i} \\\hspace*{0.5em}
\left\lfloor
\begin{array}{l}
\text{- $\boldsymbol{z}^{\mathrm{\mathrm{tr}}}_{ij} =  (\boldsymbol{p}^{\mathrm{tr}}_{ij,i} + \boldsymbol{p}^{\mathrm{tr}}_{ij,j}) / 2 $ {\footnotesize $\forall (i,j) \in \mathcal{H} \times \mathcal{H} $}}
\\ 
\text{-  $\boldsymbol{z}^{\mathrm{net}}_{i} =  (\boldsymbol{p}^{\mathrm{net}}_{i,i} + \boldsymbol{p}^{\mathrm{net}}_{i,n}) / 2$ {\footnotesize $\forall i \in \mathcal{H}$}}
\end{array}
\right.\\[1em]
\text{4. Update all dual variables } \boldsymbol{\lambda}^{\mathrm{w+1}}_{ij} \text{ and } \boldsymbol{\mu}^{\mathrm{w+1}}_{i} \\\hspace*{0.5em}
\left\lfloor
\begin{array}{l}
\text{-  $\boldsymbol{\lambda}^{\mathrm{tr},w+1}_{ij,i/j} = \boldsymbol{\lambda}^{\mathrm{tr},w}_{ij,i/j} + \boldsymbol{\rho} (\boldsymbol{p}^{\mathrm{tr}}_{ij,i/j} - \boldsymbol{z}^{\mathrm{tr}}_{ij})$  {\footnotesize $\forall (i,j) \in \mathcal{H} \times \mathcal{H} $}}
\\ 
\text{-  $\boldsymbol{\mu}^{\mathrm{net},w+1}_{i,i/n} = \boldsymbol{\mu}^{\mathrm{net},w}_{i,i/n} + \boldsymbol{\rho} (\boldsymbol{p}^{\mathrm{net}}_{i,i/n} - \boldsymbol{z}^{\mathrm{net}}_{i})$  {\footnotesize $\forall i \in \mathcal{H}$} }\end{array}
\right.\\[1em]
\hspace*{1em}
\vspace*{-0.4em}
\end{array}
\right.
$\\[-0.6em]
\hspace*{1.5em} $\boldsymbol{w} \leftarrow \boldsymbol{w}+1$\\
\textbf{Sensitivity computation:} Compute $\boldsymbol{s}_{\boldsymbol{\gamma}}(\boldsymbol{p}_i)$ {\footnotesize $\forall i \in \mathcal{H}$} using \eqref{eq:ch5_final_sens}
\\[.3em]
\textbf{Return: } Trades $ \boldsymbol{p}^{\mathrm{tr},k+1}_{ij}  \leftarrow \boldsymbol{z}^{\mathrm{tr}}_{ij}$, $ \boldsymbol{p}^{\mathrm{net},k+1}_{i} \leftarrow \boldsymbol{z}^{\mathrm{net}}_{i}  $ \\
\hspace*{4.15em}Duals $\boldsymbol{\lambda}^{k+1}_{ij} \leftarrow \boldsymbol{\lambda}^{h+1}_{ij}$, $\boldsymbol{\mu}^{k+1}_{i} \leftarrow \boldsymbol{\mu}^{h+1}_{i}$\\
\hspace*{4.15em}Sensitivities $\boldsymbol{s}_{\boldsymbol{\gamma}}(\boldsymbol{p}_i^{k+1}) \ \forall i \in \mathcal{H}$\\
\hspace*{4.15em}Optimal setpoints $\boldsymbol{p}^*$ and $\Gamma^*$
\medskip
\hrule
\end{figure}

\vspace{-0.3cm}
\subsection{Fair Bilateral trading prices}
\label{sec:ch5_fairpricing_P2P_loss}
Once the optimal dispatch of the hubs and the corresponding trade values are set, the computation of the bilateral trading prices  is done using the strategy proposed in \cite{BEHRUNANI20233751}. This involves solving an optimization problem with a constraint set of the form
\begin{equation}\nonumber
 \mathscr{C} :=\left \{\boldsymbol{c}~\begin{array}{|l} J_i(\boldsymbol{p}_i^{\star},\boldsymbol{c}_i) \leq  J_i^{\text{nt}}, \ \forall i \in \mathcal{H}, \\ \boldsymbol{c}_{ij}=\boldsymbol{c}_{ji}\ \forall (i,j) \in \mathcal{H} \times \mathcal{H} \text{, other constraints}
 \end{array} \right \}
 \end{equation}
where $J_i^{\text{nt}}$ is the cost incurred by the hubs without P2P trading and $J_i( p_i^{\star},c_i)$ is the actual cost of the hub that can be computing by calculating the objective of \eqref{eq:ch5_energyhub_optimization} using the optimal solution $p_i^{\star}$ and $c$. 

The first constraint ensures that bilateral trading prices are locally beneficial to all hubs, that is, the prices are set in such a way that the cost of each hub does not exceed the cost without P2P trading. The second constraint ensures that the trade price is the same for both hubs. Feasibility of these two constraints is shown in \cite{BEHRUNANI20233751}. The constraint set can also include constraints such as price caps and other regulations.  In \cite{BEHRUNANI20233751} fairness is defined as minimizing the variance of the normalized cost reduction defined as
\begin{align}
\nonumber
\boldsymbol{d}_i\left(\boldsymbol{c}_i\right) = \frac{(J_i^{\text{nt}}-J_i\left(\boldsymbol{p}_i^{\star},\boldsymbol{c}_i\right))}{J_i^{\text{nt}}},
\end{align}
among the hubs. This gives rise to the optimisation problem 
\begin{equation}%
\label{eq:ch5_hub_cost_balance}
\min_{\boldsymbol{c}} \ \ \  \sum_{i\in \mathcal{H}}\left(\boldsymbol{d}_i(\boldsymbol{c}_i)-
\frac{1}{H}\sum_{i\in \mathcal{H}}\left(\boldsymbol{d}_i(\boldsymbol{c}_i)\right) \right)^2 \text { s.t. } ~ \boldsymbol{c}\in \mathscr{C}.
\end{equation}

In \cite{BEHRUNANI20233751}, a semi-decetralised mediation protocol is proposed to solve \eqref{eq:ch5_hub_cost_balance}, to preserve privacy and improve scalability with respect to the number of hubs. A virtual mediator is introduced between each pair of hubs $i$ and $j$, whose objective is to determine a fair trading $\boldsymbol{c}_{ij}$. The mediator receives the cost reductions $d_{i}(\boldsymbol{c}_{ij}^k)$ from the hubs $i$ and $j$ and updates the price according to a projected-gradient descent step,
\begin{equation}
\label{eq:ch5_grad_descent}
\boldsymbol{c}_{ij}^{k+1}=\mathbb{P}_{\mathscr{C}}\left[ \boldsymbol{c}_{ij}^k - \beta_{1} \nabla_{\boldsymbol{c}_{ij}}  \varphi(\boldsymbol{p}^*,\boldsymbol{c}^k)\right]
\end{equation} 
and communicates it back to the hubs. The hubs then report their new cost reduction and the process is repeated. To compute $\nabla_{\boldsymbol{c}_{ij}}  \varphi(\boldsymbol{p}^*,\boldsymbol{c}^k)$, the mediators require the average cost reduction of all the hubs in the network. This can be computed either by communication between the mediators or through a central coordinator that receives the cost reductions from each mediator and computes the average that is then communicated back to the mediators. Details of the complete algorithm and the proofs can be found in \cite{BEHRUNANI20233751}.

The complete optimization framework for the loss-aware pricing strategy and trading price computation in summarized in Alg. \ref{alg:Optimization_pseudocode}. The optimal operation of the hubs and the network over the horizon $\mathcal{T}$ is determined at intervals of $T$ time steps. During the interim, the system only applies the previously computed optimal control inputs. To speed up convergence, the algorithm is warm-started using the optimal trade and dual values from the previous time step. To reduce computational load, trade prices can be updated every $T_f$ time steps for the trades that occurred in the past $T_f$ time steps, where, for simplicity, we assume that $T_f$ is an integer multiple of $T$.

\begin{figure}[t] \label{alg:Optimization_pseudocode}
\flushleft
\hrule
\smallskip
\textsc{Algorithm IV}: Optimization pseudo-code
\smallskip
\hrule
\medskip
\textbf{Initialization: } $t=0$\\ 
\textbf{while} true \textbf{do:}\\ [.3em]
$
\left\lfloor
\begin{array}{l l}
& \text{if }t \ \% \ T = 0 { :}\\
& \hspace*{-1em}\left\lfloor
\begin{array}{l l}
 & \hspace*{-1em} \text{1. Measure the current state}\\
& \hspace*{-1em} \text{2. Compute the setpoints/losses when no trading occurs}\\
& \hspace*{-1em} \text{3. Compute the optimal trades and tariffs for the}\\
&  \text{     complete horizon $T$ using Alg. \ref{alg1:ADMM_hypergrad}}\\
& \hspace*{-1em} \text{4. Settle the tariff payment between hubs and network}\\
& \hspace*{-1em} \text{5. If }t \ \% \ T_f = 0 { :}\\
&  \hspace*{-1em}\left\lfloor \begin{array}{l l}
 & \hspace*{-1em} \text{a. Compute bilateral trading prices for the trades} \\
 & \text{of the past $T_f$ hours using \eqref{eq:ch5_hub_cost_balance}, \eqref{eq:ch5_grad_descent}}\\ 
 & \hspace*{-1em} \text{b. Settle the payment for trading between hubs}
\end{array}
\right.
\end{array}
\right. \vspace{0.3em} \\ \vspace{0.2em}
& \hspace*{-1em} \text{Apply the optimal control input for time } t \\
& \hspace*{-1em} t \leftarrow t+1
\end{array}
\right.
$
\medskip
\hrule
\end{figure}

\section{Numerical Experiments}


We perform an extensive numerical study on the IEEE 33 bus benchmark shown in Fig. \ref{fig:ch5_elect_grid_hub}(a)~\cite{9647175}. This test system, with radial topology, has 33 buses, 32 branches that connect the bsses, and one feeder substation of 12.66 kV. The voltage limits are $\bar{\boldsymbol{v}} = 1.05$ p.u. and $\underline{\boldsymbol{v}} = 1.05$ p.u. for all buses and the phase angle is limited to $\underline{\boldsymbol{\theta}} = -0.75$ and $\bar{\boldsymbol{\theta}} = 0.75$. The network is simulated  with five hubs connected to buses 4, 7, 10, 20, 23. To test scalability, the network is extended to include up to 20 hubs, each linked to a different bus in the network. The demand profiles of the hubs are obtained from buildings at the ETH and Empa campus in Switzerland. The energy hub for each building is designed based on either the real devices and capacities present therein or by designing new hub configurations. We assume that each hub in the considered system has a perfect forecast of the daily profile of its load demand, local renewable generation and of the energy prices. Details of each hub, all data sets and the complete code is available on Gitlab\footnote{https://gitlab.ethz.ch/bvarsha/loss-aware-pricing}. Hub 1 represents a larger industrial hub with a high energy production capacity, Hub 2, 4 and 5 are medium sized hubs, and Hub 3 is a small residential hub. The electrical energy prices are time varying based on the peak hours whereas the gas prices are fixed throughout the day. The price of the electricity, $\boldsymbol{c}^{\mathrm{out}}_{\mathrm{e}}$, during peak/off-peak hours is 0.27/0.22 CHF/kWh, the feed-price of electricity, $\boldsymbol{c}^{\mathrm{in}}_{\mathrm{e}}$, is 0.12 CHF/kWh and price of gas, $\boldsymbol{c}^{\mathrm{out}}_{\mathrm{g}}$, is 0.115 CHF/l based on the Swiss market. In this study, we set $\boldsymbol{\gamma}_{ij}$ to be constant over the complete horizon $\mathcal{T}$ in order to reduce the computational complexity and speed up convergence. This also aligns with present market practices where such values are typically constant. We use Gurobi in Python, on a laptop with Windows 10, an Intel Core i7-8565U CPU @ 1.80GHz, 4 cores, 32 GB RAM.

The simulations are conducted for a period of 9 days in December 2018. We use a sampling time $t = 1$h, and a horizon length $T = 24h$ and settle P2P trades once for the 9 days, $T_f = \SI{24}{\hour} \cdot 9 = 216\SI{24}{\hour}$. The tolerance for the inner loop $\boldsymbol{\epsilon}$ is set to 0.2 and the maximum number of iterations allowed, $\bar{\boldsymbol{w}}$, is set to 100, following \cite{BEHRUNANI2024105922}. The tolerance for the outer loop $\boldsymbol{\sigma}$ is set to 0.2 and the maximum number of iterations allowed for the distributed hyper-gradient descent algorithm, $\bar{\boldsymbol{k}}$, is set to 30. The step sizes for the outer loop $\left\{\boldsymbol{\alpha}^k, \boldsymbol{\beta}^k\right\}_{k \in \mathbb{N}}$ are set to $2e\!-\!6$ $\cdot$ $0.1^{\lfloor\frac{k}{10}\rfloor}$ and 1, respectively, in accordance to the conditions set in \cite{big_hype} and based on empirical evaluation. These parameters were found to provide a good trade-off between computation and performance. The end of the simulation window is also treated in the same way as the rest of the simulation and has a prediction interval of \SI{24}{\hour} into the future. In our study, the maximum iterations is never reached and the algorithm always converges by being below the tolerance.

\vspace{-0.3cm}
\subsection{Impact of loss aware trading tariffs}
\label{results1}

\begin{figure}
\centering
\includegraphics[width=89mm]{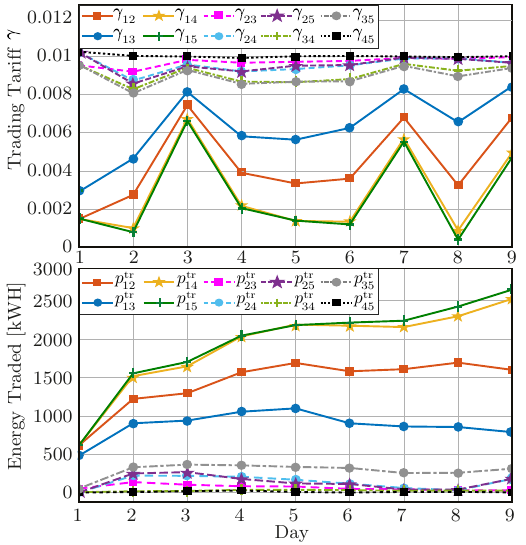}
  \caption{Results of the proposed algorithm: (a)Trading tariff computed for every trade in the network each day and (b) the resulting energy traded between the hubs in the network.}
  \label{fig:ch5_gamma_trades}
 \end{figure} 

\begin{figure}
\centering
\includegraphics[width=89mm]{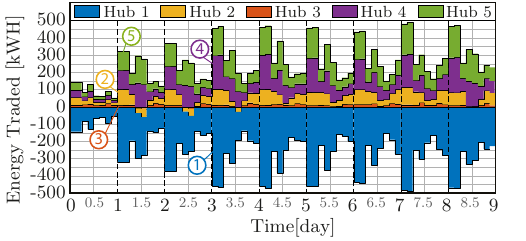}
  \caption{Net energy traded by each hub in the network.}
  \label{fig:ch5_nethub_trade}
 \end{figure} 
Fig. \ref{fig:ch5_gamma_trades} shows the optimal trading tariff computed by Alg. \ref{alg1:ADMM_hypergrad} for each day and every trade in the network. The results show that the $\boldsymbol{\gamma}_{ij}$ values are lower for hubs that are closer to one another, as well as along lines in the network that generate lower losses. Fig. \ref{fig:ch5_gamma_trades} also shows the resulting energy traded between the energy hubs in response to these tariffs. As expected, more energy is traded between hubs that have a lower trading tariff. Since Hub 1 has a large production capacity and is centrally located, the trading tariffs with Hub 1 are lower than those of others, as trading with Hub 1 reduces the losses in the network and fulfills electricity demand at a lower cost. The lowest trading tariffs are between Hub 1 and 4 and Hub 1 and 5. The tariff between Hub 1 and Hub 3 is slightly higher, despite Hub 3 being closer to Hub 1. This is because Hub 3 is a much smaller hub with significantly smaller energy demand, and a higher tariff is needed to sufficiently recover the losses. Hub pairs such as (4,5) and (2,3), which are located far away from each other and are both consuming hubs, are discouraged from trading by a higher trading tariff, resulting in negligible trade between them. Fig. \ref{fig:ch5_nethub_trade} illustrates the net energy traded by each hub. The net energy traded is negative for a producing hub and positive for a consuming hub. Electrical energy is traded mostly from Hub 1 to the other hubs, as it is the largest and has higher production capacities. This transfer is also possible due to multi-generation units, such as CHP, that allow for co-generation of electricity and heat at a much lower cost than purchasing electricity from the grid. Additionally, the central location of Hub 1 results in lower losses compared to  importing electricity from the main grid. 

 \begin{figure}
\centering
\includegraphics[width=89mm]{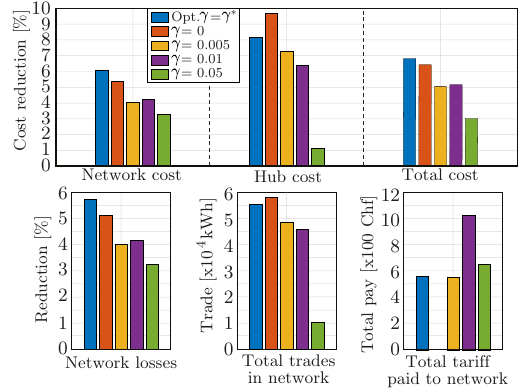}
  \caption{Overall cost reduction, reduction in network losses, total trades and the total trading tariff paid to the network using optimal $\boldsymbol{\gamma}^{\star}$ computed using the proposed algorithm and constant $\boldsymbol{\gamma}$ values of 0, 0.005, 0.01, and 0.05.}
  \label{fig:ch5_compare_other_gamma}
 \end{figure} 
 
Fig. \ref{fig:ch5_compare_other_gamma} compares the results obtained using the optimal $\boldsymbol{\gamma}^{\star}$ with those using constant grid tariff values of $\boldsymbol{\gamma} = 0, 0.005, 0.01, 0.05$ for all trades. For these values, the problem is solved using only the distributed consensus ADMM presented in Alg. \ref{alg:admm_sens} with the specified $\boldsymbol{\gamma}$. The cost reduction is computed relative to the cost when no P2P trading occurs in the network, and each hub can only exchange energy with the electricity grid. It is evident that using the optimal $\boldsymbol{\gamma}^{\star}$ results in the greatest cost reduction for the network and the total system cost. The total system cost is defined as the sum of the network cost and the operational costs of all hubs in the network. Using $\boldsymbol{\gamma} = 0$ achieves a higher cost reduction for the hubs and results in higher trade volumes since hubs do not incur any trading penalties. However, this comes at the expense of a increased network cost due to increased network losses. Consequently, the overall system cost is higher, as the lower hub costs are insufficient to offset the increase in network costs. The optimal $\boldsymbol{\gamma}^{\star}$ also leads to a greater reduction in network losses. While the algorithm is designed to recover the cost of additional network losses, it simultaneously incentivizes trades that reduce losses. Although the total trading tariff paid to the distribution system operator (DSO) is nearly the same for $\boldsymbol{\gamma}^{\star}$ and constant $\boldsymbol{\gamma} = 0.005$, the total trades are higher, network losses, and overall costs are significantly lower when using the optimal $\boldsymbol{\gamma}^{\star}$. Conversely, using higher values of $\boldsymbol{\gamma} = 0.01$ and $\boldsymbol{\gamma} = 0.05$ substantially reduces the benefits while increasing the payment to the DSO. For $\boldsymbol{\gamma} = 0.05$, the trades are reduced to less than 20\% of the trades achieved with the optimal $\boldsymbol{\gamma}^{\star}$, resulting in minimal cost reduction for the hubs. The operational cost in this case remains nearly the same as when no trading occurs, despite the total tariff paid to the network being higher than when trading is five times greater. The total tariff is highest for $\boldsymbol{\gamma} = 0.01$ because the trade volume is comparable to that with $\boldsymbol{\gamma} = 0.05$, but the trades are penalized twice as heavily. This comparison highlights the advantages of using an optimal trading tariff that is computed individually for each pair of hubs based on network topology, losses, and hub costs, rather than applying a constant tariff uniformly to all trades.  

  \begin{figure}
\centering
\includegraphics[width=89mm]{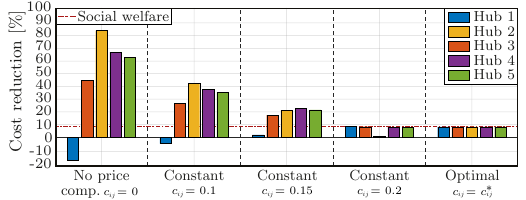}
  \caption{Cost reduction for different trading prices and the optimized prices. The social welfare cost reduction of the network is also depicted and is independent of the bilateral trading price. }
  \label{fig:ch5_hub_cost_red}
 \end{figure} 

\vspace{-0.3cm}
\subsection{Impact of trading prices}

Fig. \ref{fig:ch5_hub_cost_red} shows the cost reduction achieved by each hub for three different constant trading prices, namely, $\boldsymbol{c} = 0.1, 0.15, 0.2$ CHF/kWh and the trading price computed using the distributed implementation of \eqref{eq:ch5_hub_cost_balance}. The figure illustrates the reduction in social welfare cost, measured as the decrease in the total cumulative cost of all hubs compared to a scenario without trading. Trading benefits the hubs by reducing costs by approximately 8.2\% (as shown in Fig. \ref{fig:ch5_compare_other_gamma}), and this cost reduction remains unaffected by the bilateral trading price. In the first two scenarios, when no trading price is computed ($\boldsymbol{c} = 0$) and $\boldsymbol{c} = 0.1$, the importing Hubs~2,~3,~4, and ~5 benefit by trading as they import cheaper energy from Hub~1. However, this results in an increase of the cost for Hub~1 since the trading price is too low (even lower than the feed-in tariff) and does not cover the additional production costs of the power traded to the other hubs. 
For $\boldsymbol{c} = 0.15$ CHF/kWh, although each of the hubs benefits from trading, the cost reduction varies drastically between the hubs. Hub~1 that exports much of its energy has a much smaller benefit relative to its total size and total cost than other hubs that only import energy. If the trading price is increased further to $\boldsymbol{c} = 0.2$ CHF/kWh, all hubs continue to benefit, however, the benefit for the small Hub~3 diminishes since the higher trading price for import and the trading tariff brings the net price of the traded energy close to the grid price. 
Finally, the pricing mechanism that minimizes the variance in the normalized cost reduction, as outlined in Section \ref{sec:ch5_fairpricing_P2P_loss}, ensures that each hub achieves a nearly equal normalized cost reduction, matching the overall social cost reduction of the network. The figure illustrates the undesired impact of P2P trading, and the need to design fair bilateral prices that incentivize all hubs in participating in trading by sharing the benefits. Minimizing the variance of hub cost reductions sets the trading price to make each hub's cost reduction as close as possible to the reduction in social welfare cost.

\vspace{-0.3cm}
\subsection{Adaptability and impact of network, hub configuration, and initialization}

In this section, we analyze the benefits and performance of the proposed algorithm under different network topology, weather conditions, hub configurations, and algorithm initialisation. The different simulation scenarios are described in Table \ref{tab:ch5_scenarios}; All other details of the network setup, hub configurations and
simulation remain the same. Furthermore, all scenarios except ``Init 0.03" are initialized with $\boldsymbol{\gamma}_{ij}^0 = 0.01$. 

Fig. \ref{fig:ch5_compare_all} illustrates the results of the simulation under the different scenarios. It shows the reduction in the network losses, the hub cost and the network cost for each scenario compared to when no trading occurs under the same conditions. Additionally, it provides the proportion of the total hub cost that is paid to the network. 
\setlength{\tabcolsep}{3pt}
 \begin{table}[h]
        \centering
    \centering
    \begin{tabular}{p{1cm} p{7.4cm}}
                \hline
                Scenario & Description \\ \hline 
                Init 0.01 &  Original network setup initialized with $\boldsymbol{\gamma}^0 = 0.01$ as in Sec.\ref{results1}.\\
                Init 0.03 & Same network setup as above initialized with $\boldsymbol{\gamma}^0 = 0.03$. \\ 
                March & Original network setup simulated for 9 days in March 2018.\\
                June & Original network setup simulated for 9 days in June 2018.\\ 
                Sept. & Original network setup simulated for 9 days in September 2018.\\
                Config 1 & Hubs 2 \& 3 connected to bus 33 \& 18, respectively.\\
                Config 2 & Hub 1 connected to bus 30 instead of bus 4. \\ 
                Alt. hub & Original network setup with different demands and hub config.\\
        \end{tabular}
        \vspace*{0.1cm}
        \caption{Different simulation scenarios.}
                \label{tab:ch5_scenarios}  
                \vspace*{-0.4cm}
\end{table}

Comparing the results of scenario ``Init 0.03" to the original ``Init 0.01" shows that the initialization has a small impact on the results. The network losses and the network cost reduction are nearly equal, however, the proportion of total hub cost paid to the DSO as network tariff for trading doubles resulting in a higher cost for the hubs. The total paid to the network accounts for more than 2\% of the total hub cost when the algorithm is initialized with $\boldsymbol{\gamma}_{ij}^0 = 0.03$ compared to just $\sim$1\% for the nominal case. The difference arises because the hyper-gradient descent algorithm converges to a local equilibrium with much higher values of optimal gamma. Nevertheless, in both cases all the network constraints are met and the trades are optimal and always recovers the cost of the losses.

The results of the simulations in ``March", ``June", and ``Sept." demonstrate that our method performs well compared to when no trading occurs under different seasonal conditions. The lowest network losses and network cost are obtained during December and June while the benefits are much lower during March and September. This is because of the increase in P2P trading during the peak winter and summer months. During winter, this is due to the higher thermal demand which results in a higher electricity demand and an increase in local production via cogeneration units and P2P trades. During the summer, this is largely due to the higher PV production. The increase in PV production in June and September is also a key factor for the significant reduction in the hub cost and excess production is used for trading with the other hubs to reduce the overall social cost of the hubs. The proportion of hub cost paid as grid tariff remains largely unchanged over the year and remains lower than 1\%. 
 \begin{figure}
\centering
\includegraphics[width=89mm]{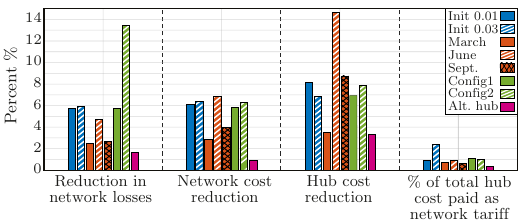}
  \caption{Simulation results under different scenarios}
  \label{fig:ch5_compare_all}
 \end{figure}

The results of ``Config 1" and ``Config 2" show that the algorithm also results in reduction in costs and network line losses under different network connections. Moving the consuming hubs (Config 1) results in nearly the same network losses and cost reduction as the original configuration (Init 0.01) and has an overall negligible impact. The relocated hubs are still able to trade with Hub 1, albeit at a fractionally higher trading cost. In contrast, changing the location of the producing hub (Config 2) significantly decreases the network losses and, to a lesser extent, hub cost. This is because, in ``Config 2", trades between hubs are along lines that have much lower losses. Although the total trade in the network decreases (specifically between Hub 1 and 2, and Hub 1 and 5) the proportion of the hub cost paid to the grid for these trades remains almost the same due to the larger trading tariff between these hubs. 

In the ``Alt. hub" scenario, the original five hubs are replaced with alternative hubs featuring different technologies, capacities, and demand profiles, while remaining at the same locations to analyze the impact of hub type on costs and losses. The 5 hubs used in this case are all medium sized commercial hubs. This results in higher network losses, network cost and overall costs. Overall, the benefits of trading are lower in this case as there is no large industrial hub, resulting in lower trading between hubs. As a consequence, the total amount paid to the grid in network tariff is also significantly reduced in this case. 

In addition to ``Config 1", ``Config 2" and ``Alt. hub", 20 other configurations were simulated with different hubs and locations. The results are not included in the interest of space, but show that the proposed method resulted in reduced network losses, network costs and hub costs in all cases, though the amount varied substantially. On average, the reduction in the network losses, network costs and hub costs was 4.2\%, 3.8\% and 6.3\% respectively, and the proportion of total hub cost paid top the network for using the grid for P2P energy trades was 0.9\% compared to no trading. These results demonstrate that the algorithm operates effectively and benefits both the hubs and the network, for a wide range of hub types, locations and seasonal variations, while ensuring that the cost of network losses are distributed fairly among network participants.

\vspace{-0.3cm}
\subsection{Scalability and computation time}

\begin{figure}
\centering
\includegraphics[width=89mm]{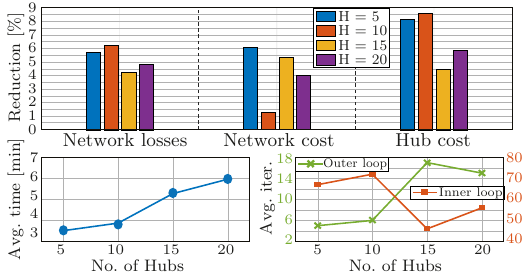}
  \caption{Results with different number of hubs, $H$, in the network }
  \label{fig:ch5_compare_all}
 \end{figure} 
 
Fig. \ref{fig:ch5_compare_all} summarises the results when the number of hubs is increases to 10, 15 and 20 hubs. Each hub is unique and positioned at specific network nodes. Their locations and details are available on GitLab. The results demonstrate that P2P trading improves losses in the network, the network cost and the operational costs of the hubs in all cases. The figure also depicts the average time and the average number of iterations of the inner and the outer loop in each case. The average time represents the time taken by the hyper-gradient descent algorithm to converge, averaged over 9 days. Similarly, the average number of iterations for both the outer and inner loops is calculated by averaging the iterations required for the algorithm to converge, also over a 9-day period. The average time required increases as the network size grows, with the lowest average time observed being 3.5 minutes for a network of 5 hubs. However, the time increase from a network of 5 hubs to 10 hubs is quite small, and there is a stronger increase for networks with 15 and 20 hubs, which take 5–6 minutes to converge. This is in part due to running the complete simulation on a single computer, resulting in a large memory load. In reality, parallel simulations would run on different energy hubs and the network, and the computational power would not be shared between them, which would significantly improve computational performance. The number of iterations for the inner loop is unrelated to the number of hubs while the number of iterations of the outer loop grows, from an average of 4.5 for smaller networks to $\sim$17 for larger networks. Even though the total number of iterations in the inner loop decreases significantly when the number of hubs increases from 10 to 15, the total time increases. This is because each iteration requires more time and computational effort. It is clear that the algorithm converges in each case within a reasonable number of iterations and successfully solves the economic dispatch and grid tariff computation problem for a 24~h horizon, even for larger networks, in approximately 5 minutes when performed on a single computer. This performance is expected to improve when computation is parallelized in real-time implementations.

\vspace{-0.2cm}
\section{Conclusion}

P2P trading in decentralized energy networks poses challenges for network operators in maintaining reliable grid operation, accounting for losses, redistributing costs equitably, and incentivizing market participation. This paper introduces a loss-aware pricing strategy for P2P energy markets to design network tariffs and trading prices that minimize losses, ensure cost recovery, and fairly allocate trading costs. Using a hierarchical Stackelberg game model, we propose an ADMM-based hyper-gradient descent solution method that ensures scalability and privacy for large-scale systems. The mechanism uses location-based, dynamic tariffs to discourage trades causing excessive losses while promoting participation. Numerical simulations on the IEEE 33-bus system validate its effectiveness, achieving significant cost reductions compared to fixed tariff schemes. The framework adapts to varying network configurations, demand profiles, and seasonal conditions, demonstrating robustness and practical applicability. Future work includes extending the model to incorporate renewable variability, dynamic demand response, and thermal trading. Additionally, leveraging data-driven methods can simplify modeling and improve plug-and-play capabilities for real-world deployment.
\vspace{-0.3cm}
\bibliographystyle{IEEEtran}
\bibliography{bibliography_main}
\end{document}